\documentclass[aps,showkeys,onecolumn,notitlepage,12pt]{revtex4}
\RequirePackage{lineno}
\usepackage{graphicx}
\usepackage{endnotes}
\usepackage{amsmath}
\usepackage{threeparttable}
\usepackage{mathtools}

\begin{document}
%\title {Evidence for Magnetic Monopoles in Mirror Space-Time of Chronometric Invariant General Relativity}
%\title {Evidence for Magnetic Monopoles and other Phenomena beyond 4-D space-time, and theory thereof}
\title {I: Evidence for Phenomena, including Magnetic Monopoles, Beyond 4-D Space-Time, and Theory Thereof}
\thanks{ \copyright  Copyright R J Ellis, 2021-23.}
\author {R J Ellis}
\email [\textit{Email address:} ] {r.ellis@physics.oxon.org}
\affiliation {Corpus Christi College, Oxford OX1 4JF, UK}
\date{\today}

%is a mismatch between theory and experiment, because theory can predict phenomena in other dimensions or spaces, whilst experiment is currently limited to determining objective facts in four-dimensional (4-D) space-time.  We ask are there phenomena in other dimensions/spaces, which can be observed but cannot be determined completely objectively?  

\begin{abstract}
Brittin and Gamow have used quantum theory to predict that sunlight lowers the entropy level at the earth's surface, apparently contrary to the second law of thermodynamics.  We have found evidence for this effect, and it requires new physics to explain it.  Another little known property of light, is that it makes it possible to detect magnetic monopoles.  We present evidence, from the scientific literature, that magnetic monopoles have already been observed at low energies, but they are non-objective, possibly in another space-time, and have been ignored.  Several experimentalists have shown that \textit{magnetic monopoles only exist in the presence of intense illumination} and so do not seem to exist in their own right.  Mikhailov made repeated measurements and determined that the monopole charge is quantized as predicted by Dirac ($g=ng_D, n$=1-5); ($\bar{g}$ = (0.99 $\pm$ 0.05)$g_D$). These results are reproducible, and so should be considered physically real.  We conclude that Dirac monopoles exist, however not in 4-D, but in another space-time.  In the second part we bring together two theories to explain how light can switch (via the Brittin and Gamow effect) into another space-time predicted by chronometric invariant general relativity (CIGR), to reveal phenomena there.  Light lowers the entropy level and so reverses the arrow of time into the mirror world of CIGR, where time is directed from the future to the past, and \textit{reveals} monopoles there.  We call this the photo-mirror effect.  This explains the infinite length of Dirac's string.  These results are preliminary evidence for unification. 

 %We provide further evidence for this in later papers.

%There are limitations to the current method of experimental physics.  Theories, such as string theory, predict other dimensions and even other spaces.  However, the method of experimental physics is limited to determining objective facts in 4-D space-time.   We reason here that there may be phenomena on other dimensions or in other spaces, which cannot be determined completely objectively as those in 4-D space-time.  However, if the results are reproducible so that they can be observed repeatedly, then we suggest that such phenomena should be considered physically real.  Magnetic monopoles are an example of this, because experiment shows that they only exist in the presence of intense illumination and so are not objective.  But they are reproducible.   Mikhailov made repeated measurements of these and determined the monopole charge to be (0.99 $\pm$ 0.05)g$_D$ .  We conclude that Dirac monopoles exist, but not in 4-D space-time.
\end{abstract}

\maketitle

\section{Introduction}
Brittin and Gamow have used the quantum theory of radiation to derive an equation which predicts that sunlight shining on the earth's surface, lowers the entropy level there, apparently contrary to the second law of thermodynamics - see equation (1) below.\cite{Gamow61}  As we investigated this further, we confirmed the violation of the second law.  To explain this, we have found new physics which helps penetrate a number of unsolved problems in quantum and particle physics, such as magnetic monopoles, quark confinement, and possibly unification.  However, there are several barriers blocking progress.  We start with the theoretical barriers.

%To test this, we have done a small experiment which not only detected this reduction in entropy, but also that the effect persists, which is contrary to the second law.\cite{Ellis1}  Another example of a peculiar effect produced by visible light is the existence of magnetic monopoles when strongly illuminated.  We develop a theoretical explanation for the persistence of the effects of sunlight and the observation of magnetic monopoles, by combining the Brittin and Gamow effect with a version of general relativity. 

1.	Murray Gell-Mann said at the Conference in Honour of his 80th birthday ``I should like to emphasize particularly... the need to go against certain received ideas.  Sometimes they are taken for granted all over the world...  Often they have a negative character and they amount to prohibitions of thinking along certain lines... Now and then, however, the only way to make progress is to defy one of these prohibitions that are uncritically accepted without good reason.''\cite{Gell-Mann}  Such prohibitions often concern problems from the past.  So it follows, contrary to the current view that references should be up-to-date, that some of the references below, are old ones.  For example, another peculiar effect of light is the detection of magnetic monopoles only when strongly illuminated, in 1930.\cite{Ehrenhaft1} %Later papers challenge other assumptions outlined below.

2.	Secondly, theory is sometimes biased against experiment.  True, it is accepted that experiment is the final arbiter of reality.  However, important discoveries often get ignored, if the correct theoretical interpretation is not given.  For example, parity violation was first observed in 1928, but was rejected as an ``instrumental effect''.\cite{Rubbia}  In 1956 Lee and Yang suggested it could be violated theoretically, and Mme Wu ``discovered'' it shortly after that.   Another example is that Ir\`ene Curie and Fr\'ed\'eric Joliot failed to discover the neutron because they did not believe Rutherford's neutron hypothesis.  Chadwick realised that their January 18th 1932 results were not due to photons but evidence for neutrons, and so made the discovery a few months later.  (The Joliot-Curies also failed to discover the positron, even though they had data for it before Anderson.)  A fourth example is that the cosmic microwave background radiation from the Big Bang was first observed by A. McKellar in 1941, but misinterpreted.\cite{McKellar}  The CMB was rediscovered at the Pulkovo Observatory by Soviet scientist T.A. Shmaonov in 1957 and published in his thesis, where he determined the temperature to be 4$\pm$3$^\circ$K, but it was ignored.\cite{Shmaonov}  Finally in 1964, Penzias and Wilson detected it a third time, and showed the results to Dicke at Princeton, who realised that this was the afterglow of the Big Bang.  Finally the discovery was made.

Another example is the case of Felix Ehrenhaft who had the misfortune to make two such discoveries, firstly of fractional electric charges in 1910 onwards, and then magnetic monopoles in 1930, and get rejected for theoretical reasons twice!  We go into magnetic monopoles in more detail below.  %His work on fractional electric charges is presented in a later paper.  %There is a subtle effect which we introduce below (based upon quantum mechanics and general relativity, so it is a unified effect), which explains his results.  

%In view of the importance of linking unexpected experimental results to correct theoretical interpretation, we present the theoretical framework for our experiments here.

%the present author, who has done some simple experiments which nobody understands, has decided to develop his own theory, which is presented in this paper.  His approach is to extend little-known theories to explain peculiar experimental results (eg Ehrenhaft's), and then apply these to his own experiments. 

%Einstein once said ``Theory can tell you what you can think.''  This is fine if the theory is correct, otheriwse it can block construictive thinking.  Niels Bohr said that physicists should study philosophy.  This is the author's approach, but it then has to be brought down to earth, in mathmetaics and experoiemnt.

3.	One of the prohibitions is the second law of thermodynamics.  It is thought to be absolute, and to lead to the ``heat death'' of the Universe.  This is in effect a classical physics ``Theory of Everything''.  It is true that (superficially) there is almost overwhelming evidence that entropy tends to increase with time.  However, the Universe is a big place and we now know that baryonic matter makes up only about 4\% of the Universe.  The other 96\% consists of dark matter and dark energy; and we do not know what these are, \textit{nor what laws they obey}.  So it is illogical to assume that the second law applies to them - it may or it may not.  So it is perfectly rational to look for processes which create order out of chaos. % as we have done.  The problem is to understand the results theoretically - hence this paper.

The author has done some experiments on phenomena which apparently violate the second law of thermodynamics, and so are inexplicable.\cite{Ellis1}  It is the objective of this paper to present a phenomenological framework to understand these results.  In the process, we find that this new framework also explains experiments on magnetic monopoles, and fractional electric charges.\cite{Ellis4}  %The reasoning is mainly from experiment upwards, rather than from abstract mathematical principles downwards, although some theory is introduced.

%which the author has done, namely to detect the subtle effect just mentioned, and a new type of radiation in sunlight.  It is interesting to note that this framework explains other experiments, such as the ones on 

4.	There are also experimental barriers to solving problems in quantum and particle physics.  Firstly, particle physics has been re-branded ``high energy physics'', which is a technique, not a subject.  Low energy particle physics is still an important and active area of research.\cite{LowE}  However, it does not get the support nor attention it deserves, because of high energy physics.  High energy experiments are massive technological achievements, so low energy experiments can appear insignificant.  It is the purpose of these papers to demonstrate the reverse.  We present new approaches, both theoretical and experimental, into magnetic monopoles, quarks, preons, confinement and possibly dark matter.

5.	Furthermore, experimental physics is currently based upon determining \textit{objective} facts in 4D space-time, for example, by controlled experiment.  However, if one relies upon objective facts only, this assumes that the Universe can be reduced to objective facts, or at least if there are any non-objective aspects, they can be ignored.  There is no proof of this, and it could lead to an infinite regression.  (For example, if matter in the Universe is made from some fundamental objective substance S$_A$, then what is this made of?  Either it is something non-objective, or it is another objective substance S$_B$, and so on.)  So less-than-objective phenomena could be more fundamental than objective ones.

In order to bring experimental physics up to date and more in line with theoretical physics (which frequently incorporates other dimensions or space-times), \textit{we propose that the requirement of objectivity should be relaxed}.  For example, if one makes measurements in other spaces or dimensions then, assuming it is possible, there is inevitably some reduction in control and/or objectivity.  It is currently not recognised that such less-objective results do occur occasionally, and so they tend to be rejected because they are not objective (i.e. not in 4-D space time).  We argue that such results should be considered physically real \textit{if they can be reproduced}.  We have examined the literature and find that magnetic monopoles are an example of this.  They are only detected under intense illumination and so may be linked to the Brittin and Gamow effect.

%We present evidence below that magnetic monopoles have been discovered and rejected by the physics community, because they are not completely objective.  We present a theoretical framework to understand this, and argue that their discovery should be recognised, albeit in mirror space-time - see below. 

Our method to challenge these barriers, is to reason from experiment upwards, as opposed to that from theoretical principles downwards, because it is experiment which can guide us to the true nature of reality.  Never-the-less, we include some theory when it is available and can help us understand the experiments.

\section{Magnetic Monopoles}
We present experimental evidence from the literature, for real ($\nabla\cdot$B $\neq$ 0) magnetic monopoles, as opposed to the pseudo-monopoles ($\nabla\cdot$H $\neq$ 0) sometimes observed in spin ices or other solid-state phenomena.

Over the last 70 years there have been numerous searches for real magnetic monopoles with mostly negative results.  Compilations of these searches conclude that there is no reproducible evidence for magnetic monopoles.\cite{SurveyA, SurveyB}  But there is an assumption behind this conclusion, namely that magnetic monopoles must be particles which can be detected objectively in 4-D space-time, because that is what controlled experiment is limited too.   Firstly, in Dirac's theory there is a line connecting two monopoles which has to be \textit{infinitely long}, and yet the universe is finite.\cite{DiracA}  This infinite length of the Dirac string is normally explained away as an artefact of the calculation.  However, it is there in the theory and implies that both monopoles are outside 4-D space-time, just as the Dirac equation implies the existence of antimatter.  (In principle one could be inside 4-D space-time and the other outside, but that would require preferential treatment for one monopole over another, which the theory does not provide.  So we reject this.)  If they are outside 4D space-time, then it would not be possible to detect them objectively by the normal methods of experimental physics (e.g. by controlled experiment).  Therefore the conclusion of the above compilations is not strictly correct.  It should read ``there is no reproducible evidence for magnetic monopoles \textit{in 4-D space-time}''.  However, this is not evidence for or against magnetic monopoles because they are not predicted to be in 4-D space-time.

Furthermore, if a phenomenon is not objective, then it is currently rejected by most physicists as \textit{not} being physically real.  Therefore, the above monopole surveys usually omit most, if not all, of the references to the following experiments which provide reproducible evidence for magnetic monopoles, but of a non-objective nature.  They are non-objective because \textit{these monopoles are only visible under intense illumination}.  When the intense illumination is turned off, they disappear, in the sense that the particle being observed ceases to move as a monopole, and moves as a neutral particle or dipole.  \textit{Thus these monopoles do not seem to exist in their own right. } However, these results are reproducible, and so we argue they are physically real.  Here is a summary of the published evidence.

\subsection{Ehrenhaft}
Ehrenhaft first reported observation of single magnetic charges, which were only detectable under intense illumination, in 1930,\cite{Ehrenhaft1} before Dirac's paper in 1931.\cite{DiracA}  However, Dirac did not recognise Ehrenhaft's results.\cite{DiracB, DiracC}  Not only were Ehrenhaft's results non-objective, but they were obtained at very low energies.  So Dirac rejected them, not just because high energies imply objectiveness, but because he thought the very strong force between monopoles would require high energies to separate them.  We explain how they can be separated at low energies below.  %Curiously more physicists know of Dirac's theory than Ehrenhaft's experiments.  

Dirac's rejection of Ehrenhaft's monopoles creates another problem, namely that there would be two different types of monopole: that predicted by Dirac's theory, and that observed by Ehrenhaft.  This is unlikely.

The essence of Ehrenhaft's observations is that when microparticles of ferromagnetic substances (such as iron, nickel or cobalt) are suspended in a gas atmosphere and subjected \textit{simultaneously} to a uniform magnetic field {\it and to intense illumination by light,\/} they move as objects carrying single magnetic charges.  If the magnetic field \textbf{\textit{H}} is reversed, then the direction of motion of the single magnetic charges is reversed (magnetic dipoles would not do this).  This effect was confirmed by Benedict and Leng.\cite{Benedikt}

Ehrenhaft did a number of experiments,\cite{Ehrenhaft2} and when he did not get the recognition he felt he deserved, he made more extreme claims, such as that ``light magnetises matter''.\cite{Ehrenhaft3}  He was convinced that he had discovered free magnetic charges and should get the kind of recognition of someone such as Amp\`ere or Faraday.  He claimed he had created a magnetic current by causing the monopoles to move.\cite{Ehrenhaft4}  He also claimed to have discovered ``magnetolysis'', being the magnetic equivalent of electrolysis.\cite{Ehrenhaft5}  Many physicists were unconvinced that ''light makes magnetism``, found the effect not objectively real, and so tended to ridicule the results.\cite{Telegdi}  Einstein took the observations seriously, but wanted a better explanation.\cite{Einstein2}
   
Kemple made a review of experimental searches for monopoles up to 1961, including not only the work of Ehrenhaft, but also by his contemporaries.  He noted that other experimenters could not reproduce some of these results, and therefore concluded that this work is not evidence for magnetic monopoles.\cite{Kemple}  However, this is not strictly correct, because even though some of the experiments may not have been confirmed, the basic observation of magnetic monopoles under intense illumination, was confirmed by Benedict and Leng.\cite{Benedikt}

\subsection{Mikhailov}\index{}
There the matter might have rested, had it not been that Mikhailov repeated Ehrenhaft's magnetic charge experiment with better technique, and confirmed the result.\cite{Mikhailov1A, Mikhailov1B, Mikhailov1C}  In his first experiment, he used iron microparticles suspended in an atmosphere of argon, illuminated by a laser with power up to 1 kW/cm$^2$, and in the presence of crossed uniform electric and magnetic fields, which were switched by a square wave-form with a frequency of a few Hertz.  The particles were observed with a microscope, and moved under the influence of the crossed electric and magnetic fields (\textbf{\textit{E}} and \textbf{\textit{H}}).  By observing their motion, one could \textit{select the signs of the electric and magnetic charges of the particles being observed, thereby confirming Ehrenhaft.}

The observed microparticles had a mass \textit{M} $\leq 10^{-14} $ gram and size \textit{r} $\leq 10^{-5}$ cm, and their motion was governed by Stokes' law.  By making measurements on particles carrying both an electric and a magnetic charge, it was possible to measure the ratio \textit{g/q} independently of the Stokes' coefficient, and hence of the size of the particle.  From observations of 1200 such particles, Mikhailov found that \textit{the magnetic charge is quantized}.  But his initial value of \textit{g} disagreed with Dirac's prediction.  However, Akers pointed out that\textit{ Mikhailov had ignored components of the particle's velocity orthogonal to} \textbf{\textit{E}} and \textbf{\textit{H}}, and so this interpretation of the result could be incorrect.\cite{Akers}  

Mikhailov reanalysed his results and found that the magnetic charge in this experiment, is in fact the solution of a quadratic equation and so gives \textit{two} possible values.  One value is the one he had previously reported, the other being \textit{that predicted by Dirac}.  In order to distinguish between these two roots, Mikhailov redesigned the experiment to remove this ambiguity and also possible surface effects. 

He condensed super-saturated vapour onto solid ferromagnetic particles in a diffusion chamber, which created a smooth surface round each particle and so eliminated surface effects.  These ferromagnetic particles, surrounded by fluid, were allowed to drop through a beam of light, under the force of gravity in a magnetic field \textbf{\textit{H}}, which was periodically inverted.  Under these conditions, particles exhibiting the magnetic charge effect, fall in a zig-zag path.  He observed 428 such tracks with a mean magnetic charge of $\bar{g}$ = (2.5$_{-1.3}^{+1.6}$).10$^{-8}$ gauss.cm$^2$, which agrees with the value predicted by Dirac of \textit{g}$_D$ = 3.29 10$^{-8}$ gauss.cm$^2$ within the errors.  In this way, Mikhailov showed unambiguously that he was observing Dirac ``monopoles'', and furthermore, these were not due to surface effects on the particles.\cite{Mikhailov2}

He also re-ran his previous experiment, choosing the correct root, and found that the ferromagnetic particles carried from 1 to 5 magnetic charges.  The histogram of magnetic charges clearly shows 5 separate peaks corresponding $g = ng_D$ where n = 1 to 5, with the peaks being gaussian-like with some gaps in between.\cite{Mikhailov3}  This confirms that the magnetic charge is quantised as predicted by Dirac, and rules out Schwinger monopoles which have twice the magnetic charge ($g_S = 2g_D$).\cite{Schwinger}

The microparticles measured by Mikhailov were composite (\textit{M} $\leq 10^{-14} $ gram), so the monopoles could be composite pseudo-particles (instantons).  However, the charge of these pseudo-particles would then not be quantised with the monopole charge predicted by Dirac.\cite{Mikhailov4}

He also reanalysed his previous experiments, selecting the correct root and dividing the data by n, and obtained a narrow bell-shaped distribution  centred on $\bar{g}$ = (3.27 $\pm$ 0.16) x 10$^{-8}$ gauss.cm$^{2}$ = 0.99 \textit{g}$_D$ with an accuracy of $\pm$5\%.\cite{Mikhailov4}   Therefore, by these ingenious experiments, Mikhailov has observed Dirac monopoles, \textit{but only when illuminated by light}.  The problem is they are non-existent in their own right, because they cease to move as monopoles when the light is turned off.  There has been no satisfactory explanation for this.

\subsection{Discussion}
%In the main, these non-objective results have been rejected or ignored because they are not existent in their own right.  However they 
These results are reproducible, because several experimentalists have observed more than 1600 single magnetic charges.  Furthermore, they apparently obey gaussian statistics (eg the bell-shaped distribution) and are statistically significant.  Therefore we argue, these single magnetic charges \textit{should be considered a real physical phenomena.}  However we have shown above that surveys of the objective methods of physics have failed to detect them, and concluded there is no evidence for them in 4-D space-time.  \textit{One possible explanation is that the monopoles observed only under intense illumination, are not in 4-D space-time but in another space-time,} as predicted by Dirac's theory.

Nevertheless, this is not a complete explanation.  We also need a theory which predicts the existence of this second space-time, together with a mechanism which enables light to switch space-time into this second space-time. We now present such a combined theory.

\section{Sunlight Shining on the Earth's Surface}
We start with an existing theory of an unexpected property of light which does the switching, and then introduce a version of general relativity which predicts a more complex structure to space-time.  The basic idea is that light switches the direction of the flow of time into that of another space-time.

\subsection{Brittin and Gamow's Theory}
In a little-known theory, Brittin and Gamow have suggested that sunlight shining on the earth, pumps entropy out into space, thereby allowing negentropy to accumulate on the earth's surface.  The sun's radiation consists of high temperature photons coming from the surface at \textit{T$_{{\rm s}}$} $\simeq$ 5,900$^\circ$ K, which spreads out in space and becomes diluted.  By the time it reaches the earth's surface, it's energy density corresponds to a temperature of the earth ({\textit{T$_{{\rm e}}$} $\simeq$ 300$^\circ$ K), so these photons are not in thermodynamic equilibrium.\\
\indent Brittin and Gamow use the quantum theory of radiation to show that the net entropy change when sunlight interacts with the earth's surface is:\cite{Gamow61}
\begin{equation}
\Delta S ~ = ~ \Delta S _{s} \,\, -\,\, \Delta S _{e} ~ = ~{\frac{4}{3}} \,\, \Delta Q\,\,\, {\left (\,{\frac{1}{T}} _{s} \,\, -\,\,{\frac{1}{T}} _{e} \, \right )}
\end{equation}
which is negative because \textit{T$_{{\rm s}}$} $>$ \textit{T$_{{\rm e}}$}.  So the entropy at the earth's surface is reduced.  They reason that this is not contrary to the second law of thermodynamics because it is simply due to the temperature gradient \textit{T$_{{\rm s}}$} $>$ \textit{T$_{{\rm e}}$} $>$ \textit{T$_{{\rm space}}$}, but see below.  (Note this effect can also occur with light from an artificial source, such as an halogen lamp.)  However, there is a hidden complication, whether the source is natural or artificial.\\
\indent This mechanism enables negative entropy to build up on the earth's surface, {\it provided it can be stored.\/}  In the case of sunlight, they calculate that photosynthesis has an efficiency of about 10\% for capturing this negative entropy.  Brittin and Gamow suggest that this is the source of order for the food chain, which Schr\"odinger proposed to be a current of negative entropy.\cite{SchrodingerA, SchrodingerB}  If this is the only mechanism for storage, then this is not a purely physical theory because it relies upon plants (and hence biochemistry) to capture the negentropy.  However, we now show that there is a mechanism in physics to store the negentropy produced.

%This theory would be more interesting to physics if there is a 
\subsection{Discussion of Brittin and Gamow Effect}
%There is also a contradiction in this theory.  On the one hand, it predicts that solar photons lower the entropy level on the earth's surface, and this logically follows from the temperature gradient.  However the reduction in entropy level of a closed system is contrary to the second law of thermodynamics.  In order to remove this contradiction, we conclude that there is something missing from Brittin and Gamow's theory, or the second law of thermodynamics, or both.

%Quite a lot has been written about \textit{time} in recent decades, but mainly from a theoretical or philosophical point of view.\cite{TimeA, TimeB, TimeC}  However, we are concerned with a very specific problem. 

In classical thermodynamics, the entropy increases with the arrow of time.\cite{Eddington}  What happens to time when a solar photon interacts with the earth's surface, thereby lowering its entropy level?  Is the direction of time reversed (e.g. locally), either momentarily or more persistently, when the photon lowers the entropy level?  We conclude that it logically must be reversed, because otherwise Eddington's arrow of time would be violated, and the second law of thermodynamics also.  Therefore what is missing from Brittin and Gamow's theory, is a theory of space-time with a second time dimension which is directed from the future to the past.  (Experimental evidence for this reasoning is given in the following reference.\cite{Ellis1})

There are a number of theories with two time dimensions, but these are compactified or otherwise unsuitable.\cite{Bars1A, Vafa}  However, K\"ohn has found a solution to the cosmological problem using two time dimensions.  The second time dimension is not compactified, but it is limited to a spacial scale of the Planck length.\cite{Kohn}  Elsborg and K\"ohn have extended this theory to the problem of magnetic monopoles, and developed the theory of magnetic monopoles in this second time dimension.\cite{Kohn1}  They adopt the orthodox view that magnetic monopoles have not been observed.\cite{SurveyA,SurveyB}  Therefore they continue the assumption from K\"ohn's first paper that the second time dimension only acts on the scale of the Planck length, so that monopoles cannot be observed experimentally at the macroscopic scales now present in the Universe.  However, the above evidence for monopoles overrules this aspect of their approach, and requires the second time dimension to be macroscopic.  Furthermore, it needs to be directed from the future to the past.  Nevertheless, this an interesting paper which provides the mathematical analysis which shows that magnetic monopoles can exist in 5D (3,2) space-time.

There is, however, another theoretical approach.  There is a little-known extension of the theory of general relativity, which has a second macroscopic time dimension directed from the future to the past.

%For example, 11-dimensional extended supersymmetry in M-theory is really a 12-dimensional SUSY with an SO(10,2) symmetry.\cite{Bars1A, Bars1B, Bars1C, Bars1D}  F-theory in twelve dimensions (12-D) is similar.\cite{Vafa}  But these second time dimensions are compactified.  K\"ohn has taken a different approach and added a second time dimension to the Einstein-Friedmann equations.\cite{Kohn}  This enables him to solve the cosmological constant problem, but the second time dimension, whilst not compactified, is on the spacial scale of the Planck length.  Bars and his colleagues have done the most work in this area of two-time physics.\cite{Bars2}  However, neither of his two time dimensions help explain the phenomena which interest us.   Furthermore, there does not appear to be any GUT which predicts a second time dimension which flows from the future to the past.  

%There is, however, an extension of the theory of general relativity which predicts this.  This extension is not well known, and so we present a summary of it below.

\section{General Relativity: Chronometric Invariants}
In the 1930s, Landau and others realised that general relativity is incomplete, because it does not correct for the reference frame of the Observer.  As a result, what is observed in a specific reference frame, \textit{is not well defined by the existing theory}.  So without the Observer, general relativity is \textit{incomplete}.  The case for including the Observer is thus compelling.  Some progress was made by Landau and Lifshitz for specific cases.\cite{Landau}  Zelmanov developed the strict mathematical formalism to calculate the observable values for any tensor quantity in 1944.  However, it was not published until 1956.\cite{ZelmanovA, ZelmanovB}  The mathematical details of the theory are given in the references.  We just present a short summary of the main points here.

%One important development in general relativity was the correct introduction of the Observer by Abraham Zelmanov.  Einstein's original theory was of space-time and matter-fields, and did not include the Observer.  However, each Observer has a specific reference frame where he or she is located, for example, at particular coordinates on the earth with its own gravitational field, which is also rotating.  As a result, what is actually observed there is not well-defined by the original theory.  The values of quantities predicted by general relativity are in effect theoretical, because in practice they are not corrected for the Observer's reference frame.  

Physically observable quantities are obtained by projecting four-dimensional quantities onto the time lines and three-dimensional space of the Observer's reference frame.  \textit{Physically observable quantities must be invariant with respect to transformations of time}, and so they are {\it chronometrically invariant quantities\/}.  Thus the general case of the Observer was incorporated into general relativity in Russia in the era of the Soviet Union.  Cattaneo later obtained similar results.\cite{CattaneoA,CattaneoB,CattaneoC}   

%The importance of including the Observer in general relativity has been recognised by others.

This important extension of general relativity is not well known in the West.\cite{stringA, stringB}  Borissova and Rabounski, have developed this theory further.  They find that the chronometric invariant equations of motion for mass-bearing particles into the past and into the future, are \textit{asymmetric in time}.  They conclude there is a fundamental asymmetry of the directions of time in the in-homogeneous space-time of general relativity.  They hold up a ''mirror`` to time and find that it does not reflect completely, and that there is a different world ''beyond the mirror``.  The four-dimensional momentum vector for a particle with non-zero rest mass, $m_0$ is:
\begin{equation}
 {P^\alpha =  m_{0} }{\frac{dx^\alpha}{ds}}, ~~~{P_\alpha P^\alpha} = 1, ~~~\alpha = 0, 1, 2, 3.
\end{equation}
When a vector (or tensor) is projected onto the time line and spacial section of an observer, these projections give the physically observable quantities for that observer.\cite{ZelmanovA}  Using the properly observable time interval ($d\tau$ = $\sqrt{g_{00}}$ $dt+$ $\frac{g_{0i}}{c\sqrt{g_{00}}}$ $dx^i$),\cite{ZelmanovA, Landau} the above four-dimensional momentum vector has two projections onto the time line, namely:\cite{Borissova, Rabounski}
\begin{equation}
 \frac{P_0}{\sqrt{g_{00}}} = \pm m, ~~~~~ \text{where} ~ m = \frac{m_0}{\sqrt{1 - v^2 / c^2}}
\end{equation}
whereas it has only one spacial projection:
\begin{equation}
 P^i = \frac{m}{c}v^i = \frac{1}{c}p^i ~~~~ \text{where} ~ v^i = \frac{dx^i}{d\tau}, ~~ i=1,2,3.
\end{equation}
%They find that the four-dimensional momentum vector for a particle with non-zero rest mass, has only one projection onto the \textit{spacial} frame of the Observer, but \underline{two} projections onto the \textit{time-line} of the Observer.  
where $p^i$ is the three-dimensional observable momentum.  They conclude that any massive particle, having two time projections, \textit{exists in two observable states}, entangled to each other: the positive mass state is in our world, while the negatively charged mass state is in the mirror world.  Using the techniques of chronometric invariants, they find that there are three separate areas: our world (i.e. normal 4-D space-time), the mirror world, and a membrane which separates the two.\cite{Borissova}   

\begin{table*}[t!]
\begin{center}
Table 1: Summary of Spacial Properties of Chronometric Invariant General Relativity
\end{center}
\begin{threeparttable}
%\caption{Results of Second Experiment} %Table 2:
\begin{tabular}
{|l|l|l|l|l|l|l|} \hline
mass &\shortstack{ Particles }& \shortstack{ Energies } &\shortstack{ Class of motion } &\shortstack{ Area } &\shortstack{Time} &\shortstack{ Entropy}\\ \hline
\shortstack{m $>$ 0 }& \shortstack { massive particles} & \shortstack{ E $>$ 0} &\shortstack{move at sub-light speeds} & \shortstack{our world} &\shortstack {dt $>$ 0} &\shortstack {$\Delta$ S $>$ 0}\\ \hline
\shortstack{m = 0} & \shortstack{massless particles (photons)} & \shortstack{ E $>$ 0} & \shortstack{move at speed of light} & \shortstack{ our world} &\shortstack{ } &\shortstack { }\\ \hline
\shortstack{m = 0 }& \shortstack{ light-like vortices}  & \shortstack{ E = 0 }  & \shortstack{moving and rotating \\at speed of light} &\shortstack{the membrane} &\shortstack{dt = 0} &\shortstack { }\\ \hline
\shortstack{m = 0} & \shortstack{massless particles (photons)} & \shortstack{ E $<$ 0 } & \shortstack{move at speed of light} & \shortstack{the mirror world} &\shortstack{ } &\shortstack { }\\ \hline
\shortstack{m $<$ 0} & \shortstack{ massive particles} & \shortstack{ E $<$ 0} &\shortstack {move at sub-light speeds} &\shortstack {the mirror world} &\shortstack {dt $<$ 0} &\shortstack {$\Delta$ S $<$ 0}\\ \hline
\end{tabular}
%\begin{tablenotes}
%\end{tablenotes}
\end{threeparttable}
\end{table*}

The flow of time is well defined mathematically in general relativity.  It is determined by the sign of the derivative of the coordinate time t with respect to the proper time ($dt/d\tau$). Using $w = c^2(1-\sqrt{g_{00}}$) and $\nu_i = -c\frac{g_{0i}}{\sqrt{g_{00}}}$ Borissova and Rabounski derive the following quadratic equation: 
\begin{equation}
\left( \frac{dt}{d\tau} \right)^2 - \frac{2\nu_iv^i}{c^2\left(1 - \frac{w}{c^2}\right)}\frac{dt}{d\tau} + \frac{1}{\left(1 - \frac{w}{c^2}\right)^2} \left( \frac{1}{c^4}\nu_i\nu_kv^iv^k - 1\right) = 0
\end{equation}
the two roots of which are:\cite{Rabounski}
\begin{equation}
 \left( \frac{dt}{d\tau} \right)_{1,2} =  \frac{1}{1 - \frac{w}{c^2}} \left( \frac{1}{c^2}\nu_iv^i \pm 1\right)
\end{equation}

This equation has three possible solutions $dt/d\tau>0$, $dt/d\tau<0$, and $dt/d\tau=0$.  In our world, $dt/d\tau>0$ and time flows from the past to the future .  {\it In the mirror world $dt/d\tau<0$ and so time flows in the opposite direction.\/}  Between the two is a membrane where time has stopped $dt/d\tau=0$.  Thus the two worlds are separate, because of the membrane, but equal.  So that to an Observer (in our world), time in the mirror world flows from the future to the past.  A summary of their results is shown in table 1.\cite{Table}  

The membrane which separates the two worlds, has its own unique three-fold structure.  On our world side and the mirror world side, are streams of light-like particles (photons), moving at the speed of light, but with opposite energies and frequencies.  Between the two in the membrane, time has stopped because $dt/d\tau=0$, and so this region is a void which is purely spacial.  However, in this void there are light-like vortices, previously unknown, which have zero relativistic masses (unlike photons which, although massless, have non-zero relativistic masses).  These light-like vortices move and rotate at the speed of light, but have no energy because for them time has stopped - they are purely spacial.

In this theory, a mass-bearing particle has two time projections, one in each world, and exists in two observable states.  Each particle is in effect a four dimensional dipole object, which exists in two states: in our world with positive mass and energy; in the mirror world with negative mass and energy (NB this negative mass state is not anti-matter, because the inertial mass of anti-matter is positive).  However, they cannot ''annihilate`` or rather ''nullify`` (since the net energy is zero) because they are separated by the membrane.  Furthermore our world and the mirror world have the same background space, and \textit{the three-dimensional momentum remains positive in both sectors}.  More details are given in the references above.  

We refer to this theory of physically observable quantities, as ''Chronometric Invariant General Relativity`` or CIGR.  In CIGR, our world (4-D space-time) and the mirror world have the same background space.  So time in the mirror world is a macroscopic time dimension.  Furthermore. mirror time is directed from the future to the past, so we would expect entropy in the mirror world \textit{to be constant or decrease with our time}.

\section{Photo Mirror Hypothesis}
We make the hypothesis that light can switch matter into the mirror world state, by means of the Brittin and Gamow effect, because this reduces the entropy level which reverses the direction of time. 
\begin{equation}
 normal (x,t) ~~~~ dt/d\tau>0 ~~~~  \xrightleftharpoons[\Delta S > 0]{\Delta S < 0} ~~~~ dt/d\tau<0 ~~~~ mirror (x,-t)
\end{equation}
We predict this will occur locally where each photon interacts (in which case $\Delta Q = h\nu$ in equation 1).  This reversal could be momentary or persistent depending on the phenomenon being observed.  We call this the ''photo-mirror hypothesis``.

Note that when it occurs, this is a low energy effect for two reasons.  Firstly according to CIGR, any massive particle exists in a 4-dimensional dipole state with positive mass and energy in our world and negative mass and energy in the mirror world.  Since the mirror world state already exists, it does not require any energy to produce it.  All that is required is the reversal of the direction of time \textit{to reveal it}, which can be done by visible photons with energies of a few electron volts (equation 1).  The author provides experimental evidence for this in a separate paper.\cite{Ellis1}

The reader may question why, if photons can switch space-time into the mirror world state, it has not been observed before.  Firstly, the effect is subtle and occurs at very low energies. Secondly, physicists are so convinced that the second law of thermodynamics is absolute, that few have looked for the creation of order.  Thirdly, it switches space-time into the mirror world where phenomena are less objective and so tend to get ignored or rejected (e.g. the magnetic monopoles above).  Furthermore, any random processes which increase entropy will switch the direction of time back to normal (4-D space-time).  Limitations of this are discussed below.

%\textit{this effect has been observed} in experiments using light to reveal magnetic charges in mirror space-time, but these have been rejected or ignored, because they are at low energies and not objective.

\subsection{Explanation of Magnetic Monopoles}
The explanation for these magnetic monopoles is that photons in the intense illumination, switch the direction of time experienced by the ferromagnetic particles (via the Brittin and Gamow effect), from 4D space-time into the mirror world space-time, where the magnetic monopoles exist and can be observed.  Therefore the intense illumination does not ''make magnetism`` as Ehrenhaft claimed,  but ''reveals magnetic monopoles`` in this other space-time.

This overcomes Dirac's objection to Ehrenhaft's monopoles, namely that magnetic monopoles would only be observed at high energies, because of the very strong force between pairs of them,\cite{DiracB} in the following way.  The monopoles are in mirror space-time where the masses are negative. Therefore the attractive force between two monopoles would cause them to fly apart, so dipoles would not form.  Thus by switching the direction of time, light can reveal the monopoles at low energies.

Furthermore, Dirac also concludes that a monopole may be connected to a string extending to infinity.  If the monopoles are in one space, and the dipole is in another, then the Dirac string between a monopole and the corresponding pole of the dipole, is naturally infinitely long.  Therefore observation of monopoles in mirror space-time and of magnetic dipoles in normal 4-D space-time, provides a natural physical explanation for the infinite length of the Dirac string, and confirms this aspect of his theory.  In view of these results, Ehrenhaft, Benedict and Leng, and Mikhailov really did observe Dirac monopoles at these very low energies.

%We have shown above that the monopoles are in a separate space-time, and so not necessarily a high energy phenomenon.

%We have concluded above that under intense illumination, observation of magnetic monopoles is reproducible, so they are physically real, but do not appear to exist in 4-D space-time.  

%Furthermore, Dirac's theory provides us with a topological picture of this phenomenon.  

\section{Discussion and Limitations}
One of the problems in physics is that general relativity is the theory of the large-scale structure of the Universe (eg black holes), whereas quantum mechanics is the theory of small-scale atomic phenomena.  As a result it has not been possible to unify the two theories.  Chronometric invariants introduces a more complex structure to space-time with the second time dimension in mirror space-time being directed from the future to the past.  This enables the second law of thermodynamics to be overcome, and persistent low-entropy states to be produced.  It is perhaps not immediately obvious, but these low entropy states can be small in scale, and therefore make it easier to unify quantum mechanics and CIGR.
  
The photo-mirror hypothesis above is an example of this.  It involves both quantum mechanics (the Brittin and Gamow effect) and general relativity (CIGR), so it is a unified effect.  Therefore the theory of this will only be properly formulated by unifying quantum mechanics and general relativity (probably in the form of CIGR), and the standard model embedded within it.  Until this is done, CIGR may have certain limitations.  For example, it predicts that a massive particle is a dipole with a net mass of zero, which seems to be contrary to the evidence that more massive particles require more centre-of-mass energy to produce them.

However, mirror space-time explains two aspects of Dirac's theory of monopoles.  Firstly why they can be detected at low energies, and secondly the infinite length of the Dirac string.  This suggests that quantum mechanics should be formulated in a space-time similar if not identical to that of CIGR.  Furthermore, the author has obtained \textit{independent experimental evidence} for the photo-mirror hypothesis, which justifies its usage above to explain the magnetic monopole data.\cite{Ellis1}  In addition we have obtained separate experimental evidence for a new type of radiation in sunlight, which appears to be in mirror space-time.\cite{Ellis2}  This could be preliminary evidence that mirror space-time is the geometrical basis for the hidden sector of this potentially new unified theory.  We have also investigated the nature of matter in mirror space-time.\cite{Ellis4}  We predict that the experimental methods to detect new phenomena in this hypothetical mirror sector, will be different from those currently used in particle physics, because the effects occur at low energies, and low entropy levels (because mirror time is directed from the future to the past).

%By incorporating the Observer into general relativity, CIGR includes thermodynamics and makes predictions about smaller scale phenomena (as opposed to black holes), such as low entropy states in the mirror world with its second time dimension.  We predict that this inclusion of thermodynamics makes it possible to unify CIGR with quantum mechanics. 

%We predict that mirror space-time will became the basis for the hidden sector predicted by most grand unified theories.  We predict that any new force(s) in this mirror sector will lower entropy levels, since mirror time flows from the future to the past.

%We predict that phenomena in this second sector will be different from those in the normal sector currently studied by high energy physics.  For example, they will tend to be non-objective, collective (low entropy), low energy phenomena, rather than objective, separate particles (high entropy) currently observed at high energies.   

\section{Conclusions}
We have made the hypothesis that there may be phenomena which experiment can detect, but which are not completely objective, for example because they are not in normal 4-D space-time.  Magnetic monopoles are an example of this, because they can only be detected under intense illumination, so that when the illumination is turned off, they cease to move as monopoles, and so do not seem to exist in their own right.  However, if a phenomenon can be detected repeatedly (for example these magnetic monopoles), \textit{we suggest it should be considered physically real.}

%We have presented reproducible evidence for magnetic monopoles which appear to exist outside 4-D space-time.  We conclude that the current method of experimental physics is flawed, because it limits observations to objective phenomena in 4-D space time.  Phenomena beyond 4-D space-time, if they can be observed, are currently rejected.  The solution is to relax the criterion of objectivity, and recognise reproducible phenomena as being physically real.  This is especially the case if there is a theory for that phenomenon.

Several experimenters have observed more than 1600 magnetic monopoles under intense illumination, so they are reproducible.  They have determined that these monopoles have the charge predicted by Dirac ($\bar{g}$ = (3.27 $\pm$ 0.16) x 10$^{-8}$ gauss.cm$^{2}$ = 0.99 \textit{g}$_D$ with an accuracy of $\pm$5\%).  Furthermore, this charge is quantised ($g = ng_D$ with n = 1 to 5).  This rules out pseudo-particles (instantons) because they would not be quantised, and certainly not with the Dirac charge.

%So Dirac monopoles appear to have been observed at low energies.
%Furthermore, we have shown that these monopoles are in the mirror space time of CIGR, and this explains the infinite length of the Dirac string.  

%However, whilst this phenomenon is not objective, and therefore outside 4-D space-time, nevertheless it is reproducible and therefore should be considered physically real.  

To explain these monopoles, we combine the Brittin and Gamow effect and chronometric invariant general relativity (CIGR) to make the photo-mirror hypothesis, namely that light lowers the entropy level and reverses the direction of time, thereby switching space-time into mirror space-time of CIGR, where time is directed from the future to the past.  Therefore the photons of the intense illumination switch the ferromagnetic particles, via the photo-mirror hypothesis, into the mirror world space-time state, where the magnetic monopoles exist and are observed.  In this way, the \textit{intense illumination reveals magnetic monopoles in mirror space-time.}  

%Dirac was convinced that monopoles could only be observed at high energies, because the force between monopoles is so strong.  However, these monopoles are in mirror space-time, where the masses are negative.  Thus a strong attractive force between two monopoles will cause them to fly apart.  

We note that observation of magnetic monopoles only in mirror space-time and dipoles only in normal 4-D space-time, provides a natural physical explanation for the infinite length of the Dirac string, and confirms this aspect of his theory.  We conclude that Dirac monopoles have been observed. 

This is evidence for phenomena beyond 4-D space-time.  In effect, under certain circumstances, light gives us a window into another world. The photo-mirror hypothesis links a quantum mechanical effect (Brittin and Gamow) with general relativity (CIGR), which implies unification.   

%The author has obtained independent experimental evidence for the photo-mirror hypothesis, which will be published separately.\cite{Ellis1, Ellis2}  This justifies its use above to explain the magnetic monopole data.  

%Furthermore, magnetic monopoles are predicted by many unified theories.  Therefore evidence for these phenomena suggests unification.  

%\subsection {Limitations}
%The above theory (CIGR) has limitations, because it has not yet been properly unified with quantum mechanics, nor had the standard model embedded within it.  Therefore, some of the details and predictions of CIGR may change.  

\section{Acknowledgements}
The author thanks Dmitri Rabounski for helpful communications.

%There is no data file.

\end{document}